# Generation of adaptive cellular structures to surfaces with complex geometries using Homotopy functions and conformal transformation

A.E Gómez-Ovalle[1]

[1]Texas A&M University

**Abstract**

The optimal use of resources has motivated the engineering community to employ controlled distribution of material within their structural designs, often relying on cellular and lattice porous structures. In this research work, a computational method is implemented to generate porous lattice structures conforming to surfaces with complex geometry. This method allows solving the problem of integrating cellular structures in engineering applications with complex macro-geometries. Examples of surface transformation are shown in this report. In addition, an applied model is shown where the effects of the transformation on the mechanical properties are analyzed.

1. Introduction

Cellular materials are based on unit cells that reproduce in space, similar to the concept of unit cells in crystallography [1]. Usually, unit cells are designed in a two- or three-dimensional Euclidean space following a rectangular geometry. In this case, a unit cell is a parallelepiped, i.e., a space enclosed by three pairs of parallel surfaces. Within unit cells are the different motifs that are repeated throughout space, generating a cellular pattern. Typical motifs within unit cells correspond to beam or membrane assemblies, optimally designed to achieve unprecedented combinations of effective materials properties [2].

On a laboratory scale, the surface of freely cellular material is an arrangement of flat surface elements. These surface elements are characterized by their normals which are oriented perpendicular to the surface elements. Thus, the orientation of a particular surface element is given through the angles which its normal makes with the normals of the other surface elements. However, the surfaces that define the boundary of an engineering part are not flat, on the contrary, these are surfaces with variable curvatures [3]. Therefore, when unit cells approach the border defined by an engineering device, the normal vectors of the plane surfaces of the unit cell and the normal vectors of the surface of the engineering part do not coincide. More specifically, for every neighborhood of a point on the surface, the tangent planes are different. That is, tangent plane parametrizations do not produce parallel points. The situation mentioned above creates a challenge because, at the boundary of an engineering device, the unit cell boundaries and the engineering device do not match. So, in engineering applications, this key challenge is related to the design of cellular and architected materials while determining the best integration of a cellular pattern into the shape of the component of interest [4]. From this, , two situations occur i) there are empty spaces not filled by the unit cells or ii) the unit cells surpass the border of the engineering device.

A solution is to start from the second situation and trim the areas of the remaining unit cells to meet the outer geometry[6]. However, this leads to another issue. The motifs within the unit cells are designed to fulfill specific effective properties, which depend on the topology of the chosen motif. If this motif is incomplete, the engineering performance of the unit cell will also be incomplete, leading to end-effects and unpredicted structural performances of the structures.

Another solution is that the unit cell's geometry conforms to the engineering device boundary, ensuring that all unit cells are complete, giving a smoother end-effect condition. This type of solution creates a challenge in the modeling of cellular materials.

In order to achieve this, it is necessary to perform a transformation of the curve or surface that defines the unit cell to the curve or surface that defines the boundary of the engineering device. A mathematical concept that allows this transformation is that of *homotopy*. This concept refers to the mapping of one continuous function into another and it is achieved through the continuous deformation of one function into another [5]. Therefore, two different topological spaces are homotopic if one space can be continuously deformed in the other.

This concept has been used in graphic computing and other similar applications [6]. For example, Tai et al. applied the homotopy concept to determine the control of the shape of cross-sections between two defined contours [6,7]. In this case, the homotopy is parametrized by two scalar-valued functions that control, separately, the transition from one contour to the other and the scaling of the cross-sectional shape between the two defined contours [8]. Hence, the homotopy concept allows the transformation of the curves and surfaces that determine the unit cell to the surface of the edges of a final application piece.

This work aims to start a framework for modeling the mechanical properties of conforming cellular materials. To achieve this, continuous homotopic (algebraic topology) and conformal geometry will be implemented on unit cell design so that the initial mechanical properties are maintained. This methodology produces a functional geometric spatial gradation in the growth of the macro-structure topology. Due to the above, it is necessary to develop laws to homogenize and scale that applies to spatially heterogeneous structures. Finally, the failure mechanisms (plastic deformation and fracture) of these architecture structures will be analyzed using hybrid computational and analytical methods. In this case, the architectural structures will be elasto-plastically modeled using the Finite Element Software Abaqus®.

2. **Homotopy functions as a solution to the curve transformation problem**

**2.1. Problem Definition**

*Definition of transformation:* Figure 1 shows an example in which a geometric object (a pentagon) is adapted to another geometric object (a circle). The indicated segment's transformation represent that adapting one geometric object into another involves a transformation from one curve to another. The smooth modification specified in the definition exposes that this transformation must be step-by-step, in a continuous, non-abrupt manner. Figure 1 shows that the transformation modifies the curvature of the curve gradually.

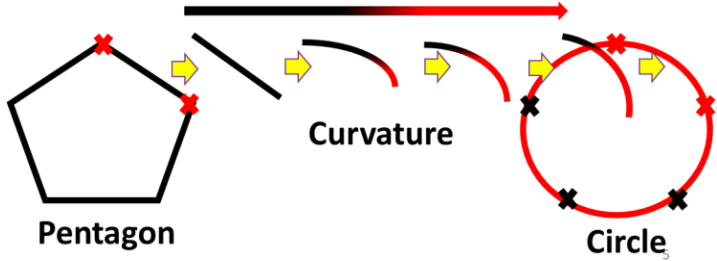

Figure. 1 Diagram of the adaptation of one structure to another. In this case, a straight line of a pentagon is transformed into an arc of a circle.

This transformation function is defined according to the concept of homotopy, which allows the geometric transformation to be carried out gradually and continuously. Homotopy is defined below:

## 2.2. Definition of homotopic function:

*Definition:* Let $f, g : XY$ be maps where $X$ and $Y$ are topological spaces. If there exists a map $F: X \times IY$ such that $F(x, o) = f(x)$ and $F(x, 1) = g(x)$ for all $x \in X$, $I = [0,1]$, then the map $F$ is called a homotopy from $f$ to $g$.

In the case of applying the definition to the pentagon problem, the function $f$ corresponds to the contour parameterized by the straight line and the function $g$ corresponds to the contour parameterized by the arc of the circle. Therefore, the homotopic function defines continuous deformation of contour $f$ into $g$ as $t$ varies from 0 to 1. The homotopic function for a contour (1D) can be expressed as follows (equation.1):

$$F(x, t) = (1 - t)f(x) - tg(x) \quad \text{eq.1}$$

In the case of a 3D surface the homotopy function is expressed as follows:

$$C(u, v) = [(1 - R_n(v))C_1(u) + R_n(v)C_2(u)] (I_2 + S(v)) \quad \text{eq.2}$$

In this equation, $C(u, v)$ corresponds to the homotopy function in generalized coordinates $u$ and $v$. $R_n(v)$ corresponds to the generalization of the parameter $t$ which is in charge of continuously transforming the initial contour $C_1(u)$ to the objective contour $C_2(u)$, this function has the name the *flexing* function. $I_2$ corresponds to the $2 \times 2$ identity matrix. The function $S(v)$, called the scaling function, determines the number of internal contours that will exist between the initial contour and the target contour.

**2.3.** Algorithm applied (pseudocode) to generate the structure:

**Algorithm 1:** Conformal transformation from an initial curve to a target curve

1: **Begin**

2: **read inputs**;

3: **if** (analytic geometry case)

4:    **Then**

5:    **Begin** {Then}

6:       determine distribution of cross-sectional planes;

7:       determine radial distribution of points;

8:       determine circumferential distribution of points;

9:       **For** each cross-section do

10:       **Begin** {each cross-section)

11:          **For** each circumferential point do

12:          **Begin** {each circumferential point}

13:       generate end-shapes;

14:        generate outer boundary;

15:        determine shape variation parameter;

16:        generate surface coordinates by blending end-shapes;

17:        **For** each radial point do

18:          **Begin** {each radial point}

19:            determine grid variation parameter;

20:            compute coefficients for orthogonal trajectories;

21:                    determine grid-point coordinates;

22:                    Write grid-point coordinates;

23:            **end;** {each radial point}

24:        **end;** (each circumferential point}

25:    **end;** {each cross-section)

26: **end'** {Then]

27: **end.**

## 3. Application of the homotopy functions algorithm to 2D cellular structures

### 3.1. The first example of application of the homotopic function (2D):

The objective in this example is to transform the contour given by the function $(y_1, z_1)$ to the contour given by the function $(y_2, z_2)$.

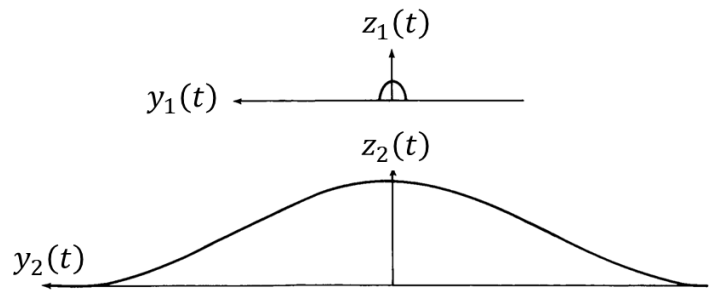

Figure.2 Contours to transform using homotopy functions.

**First step.** The first step corresponds to the parameterization of the contours. Below are the settings for both contours:

The initial shape, represented in parametric form with parameter $t$, can expressed as :

$$y_1 = y_1(t) \; z_1 = z_1(t) \}(t_a < t < t_b) \quad \text{eq.3}$$

The target cross-section in parametric form with the same $t$ domain $(t_a, t_b)$ is represented as:

$$y_2 = y_2(t) \; z_2 = z_2(t) \}(t_a < t < t_b) \quad \text{eq.4}$$

Therefore, the initial and terminal cross-section can be expressed as follows:

| Initial cross-section | Final cross-section |
|---|---|
| $y_1(t) = t$ | $y_2(t) = t$ |
| $z_1(t) = \dfrac{\sqrt{1-t^2}}{3}$ | $z_2(t) = \sqrt{(0.317)^2 - t^2}$ |

Table.1 Parameterized cross sections.

**Second Step.** In this step the bending function is determined by means of a polynomial function:

The function $c_n[x_n(x)]$ is usually specified to be smooth and monotonic. A typical expression for $c_n$ might be:

$$c_n(x_n) = x_n^m \quad \text{eq.5}$$

with *m* as an adjustable design parameter.

After that, a transition surface representing a smooth blending of the two end shapes is defined by the functions:

$$y(x,t) = [1 - c(x)]y_1(t) + c(x)y_2(t) \quad \text{eq.6}$$

$$z(x,t) = [1 - c(x)]z_1(t) + c(x)z_2(t) \quad \text{eq.7}$$

Finally, the size variation of the surface is prescribed by introducing an independent scaling function $\lambda(x)$, which is specified to be smooth and nonnegative but not necessarily monotonic:

$$y(x,t) = \lambda(x)\{[1 - c(x)]y_1(t) + c(x)y_2(t)\} \text{ eq.8}$$

$$z(x,t) = \lambda(x)\{[1 - c(x)]z_1(t) + c(x)z_2(t)\} \text{ eq.9}$$

The scaling function $\lambda(x)$ is defined by:

$$\lambda(x_n) = \lambda_2(1.6x_n - 5.3x_n^2 \quad \text{eq.10}$$
$$+9.1x_n^3 - 4.4x_n^4$$

After applying this algorithm, the determined cellular structure is generated by using the functions mentioned above:

## 4. Comparison of the mechanical behavior between a 2D conformal structure and a 2D non-conformal structure

Figure 8 shows a preliminary qualitative comparison between a non-conforming and a conforming structure. In this case, there is high deformation in the outer and adjacent unit cells. On the other hand, in the structure, it is observed that the deformation of the outer cells is less.

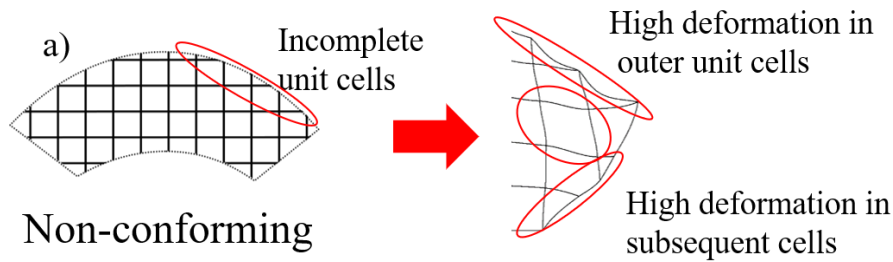

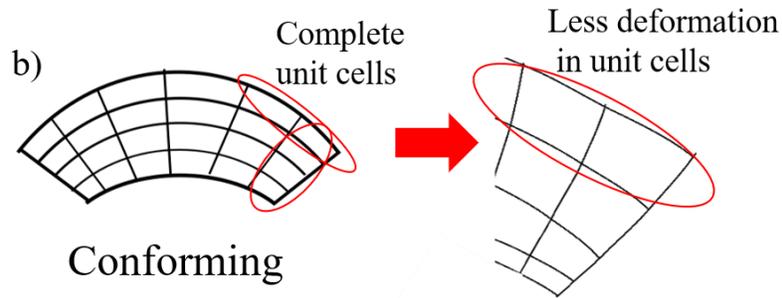

Figure. 8 Models are showing the coupling of a conformal and a non-conforming cell structure with a circular geometry edge. a) The nonconforming model shows a large deformation at the edges of the unit cells close to the edge. b) The conformal model shows a significant reduction in unit cell deformation near the edge.

The figure. 9 shows an implementation of a non-conforming structure and a conforming one in another type of geometry. In this case, a tensile load is applied at one end of the structure of 100MPa. The structure is recessed at the other end. The structure material is Titanium with a modulus of elasticity of 116GPa and a Poisson's Ratio of 0.34.

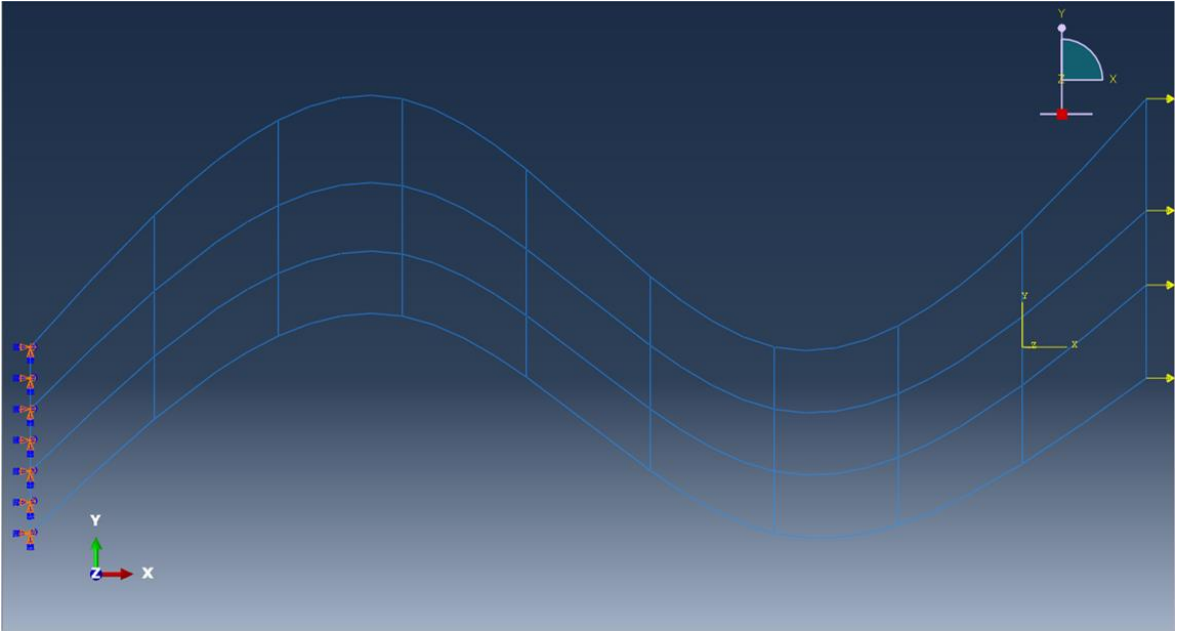

Figure.9 Finite element model of a conformal structure.

Figure.10 shows the results of the simulation. These models show three essential characteristics:

1. The non-conforming structure presents a more significant deformation along the x-axis.
2. The conforming design gives higher stress concentrations.
3. The stress distribution in the conforming system is more uniform.

This last point is observed in the red circle in the non-conforming structure. In this red circle, an area is observed in which the stresses are low compared whole system. On the other hand, the conformal structure presents a similar stress distribution throughout the design; there are no areas with minor deformation. Comparison in the stress distribution between a conforming structure and a non-conforming structure

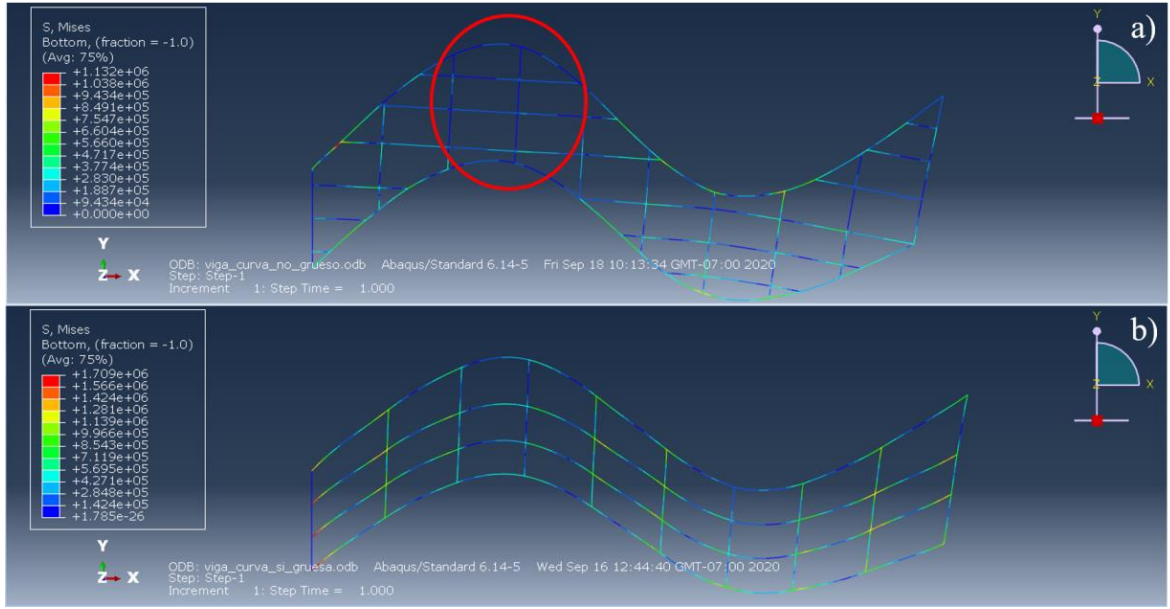

Figure.10 Comparison of the stress distribution between a conforming structure and a non-conforming structure.

From these results, the following can be concluded:

a) The conformal structure generates a homogeneous distribution throughout the entire structure. This allows achieving one of the objectives of the methodology, which is that all unit cells throughout the structure deform uniformly.

A conformal structure generates geometric alterations in the unit cell concerning the original unit cell. In this case, a pseudo-periodic structure is obtained that can decrease the mechanical rigidity of the structure. For this reason, the system changes the mechanical behavior initially expected for the original unit cells. Therefore, it is necessary to include a section in the transformation algorithm that allows controlling the mechanical behavior of the unit cell while it is transformed.

According to those mentioned above, it is necessary to generate a complement to the algorithm that allows controlling the transformation of the structure while maintaining a specific mechanical property. Therefore, the unit cell transformation modeling was divided into two categories to achieve the above objective. Figure. 11 illustrates the two ways in which unit cells are transformed.

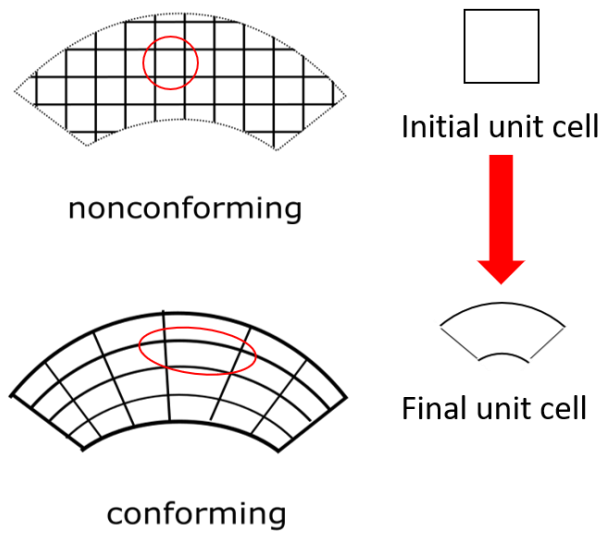

Figura.11 Two ways in which a unit cell is transformed. First, the unit cell is deformed at its angles, and second the size of the unit cell changes as it approaches the boundary.

In the first instance, the figure shows that the unit cell changes shape, more specifically the angles and edges, which are coupled to the geometry of the outer edge. Second, once the unit cell takes on a particular geometry, the size increases as it gets closer to one of the edges.

The change in shape is represented by the Jacobian operator $J$, as shown in eq. 11. On the other hand, the changes in size are represented using the function $w(x)$. the function $y$ represents any mechanical property.

$$y = Jx + \delta w(x) \quad \text{Eq.11}$$

$$J = \begin{pmatrix} \sin(\theta_1 + \theta_2) & -\cos(\theta_1 + \theta_2) \\ -\sin\theta_1 & \cos\theta_1 \end{pmatrix} \quad \text{Eq.12}$$

The next step is to generate a constitutive model of the conformal structure. For this, we use an asymptotic homogenization law. The goal is to decouple the originally multiscale problem with the use of two single-scale variables: $x$ capturing the macroscale trend and $\bar{Y}$ capturing the microscale. The two-scale spatial gradients get dissociated by

$$\frac{\partial}{\partial x_j^\varepsilon} = \frac{\partial}{\partial x_i} + \frac{J_{ik}}{\varepsilon} \frac{\partial}{\partial \bar{Y}_k} \quad \text{Eq.13}$$

Eq. 13 shows the definition of mechanical property gradient in terms of a global variable and a microscale variable. In addition, it is essential to include the $J$ term from the Jacobian.

Finally, it is possible to arrive at an expression for the energy

$$E^H \sim \frac{1}{2} \int \left[ \left( c_{ijkl}^{H[0]} - \delta c_{ijkl}^{H[1]} \right) \frac{\partial u_i^{H[0]}}{\partial x_j} \frac{\partial u_i^{H[0]}}{\partial x_l} \right] dx \quad \text{Eq.14}$$

The perspective of this work is to implement these constitutive models in the homotopic transformation algorithm and be able to control the changes in the mechanical behavior of the transformed cell structures.

5. **Conclusions**

In this work, it was possible to determine a solution methodology for implementing cellular structures in engineering devices with complex geometry. This methodology is based on the concept of homotopy by transforming the geometry of a unit cell into the geometry of an exterior cellar. This methodology is computationally efficient because it allows unit cells to be modeled as functions and not as objects in CAD software. On the other hand, a deep analysis of the transformation dynamics of the cells was made, and their limitations were evaluated from a mechanical point of view. In this case, it was determined that the transformed structure allows obtaining a more homogeneous stress distribution than a non-conforming structure.

On the other hand, it was shown that the deformation mechanisms of the conforming unit cell differ from the deformation mechanics of the structure based on the undeformed unit cells. This showed the need to optimize the algorithm because it does not consider the mechanical behavior of the system. It is necessary to build a constitutive model that models the conformal structures' mechanical behavior to perform this optimization. In this case, the transformation of the structure generates changes in the morphology of the unit cell, and to generates a size gradient as it approaches the edge of the target surface. To model these changes, we implement a property function that contains a Jacobian function and a function that describes the transformation in size. Finally, as a perspective of this work, this constitutive model must be implemented in the original algorithm to generate a homotopic transformation that allows controlling the mechanical behavior of the structure. The perspective of this work is to implement these constitutive models in the homotopic transformation algorithm and be able to control the changes in the mechanical behavior of the transformed cell structures.

## 6. References

1. Wahab MA. Unit Cell Symmetries and Their Representations. Numerical Problems in Crystallography. 2021. pp. 209–262. doi:10.1007/978-981-15-9754-1_6
2. Hanks B, Frecker M. Lattice Structure Design for Additive Manufacturing: Unit Cell Topology Optimization. Volume 2A: 45th Design Automation Conference. 2019. doi:10.1115/detc2019-97863

3. Hanks B, Frecker M. 3D Additive Lattice Topology Optimization: A Unit Cell Design Approach. Volume 11A: 46th Design Automation Conference (DAC). 2020. doi:10.1115/detc2020-22386

4. Modelling material uncertainty in multiscale optimization using lattice microstructures. 2020. doi:10.2514/6.2021-1593.vid

5. James IM. General Topology and Homotopy Theory. Springer Science & Business Media; 2012.

6. Agoston MK. Computer Graphics and Geometric Modelling: Mathematics. Springer Science & Business Media; 2005.

7. Tai C-L, Loe K-F, Kunii TL. Integrated Homotopy Sweep Technique for Computer-Aided Geometric Design. Visual Computing. 1992. pp. 583–595. doi:10.1007/978-4-431-68204-2_36

8. A Special Issue of Computer Aided Geometric Design. Computer Aided Geometric Design. 2007. p. 371. doi:10.1016/s0167-8396(07)00071-4